\documentclass[11pt,a4paper]{article}
\usepackage{jheppub,amsmath,  amssymb,slashed,url,bm,textgreek,upgreek,footmisc}
\usepackage{graphicx}
\usepackage{epstopdf}

\usepackage{tikz}
\usetikzlibrary{decorations.pathreplacing, decorations.markings,calc,shapes.misc,decorations.pathmorphing,patterns.meta, math}

\usepackage{caption}
\usepackage{subcaption}

\usepackage{booktabs}
\usepackage{diagbox}
\usepackage{array}
\usepackage{multirow}
\usepackage{braket}
\usetikzlibrary{matrix,arrows.meta}
\usepackage{comment}

\newcommand\SU{\text{SU}}

\newcommand{\beq}{\begin{equation}}
\newcommand{\eeq}{\end{equation}}
 
\newcommand{\bea}{\begin{eqnarray}}
\newcommand{\ea}{\end{eqnarray}}

\def\d{{\mathrm d}}

\def\i{{\mathrm i}} 

\def\r{{\textbf r}}

\newcommand{\tr}[1]{\textrm{Tr}[#1]}

\def\SU{{\textrm{SU}}}

\def\su{{\mathfrak{su}}}

\title{Fortuity with a Single Matrix\vspace{-0.5cm}}

\author{Yiming Chen${}^{1}$}

\affiliation{${}^1$ Leinweber Institute for Theoretical Physics at Stanford, Stanford, CA 94305, USA}

\abstract{
We construct and study a supersymmetric quantum mechanical model with a single $U(N)$‑adjoint fermionic matrix. The model is exactly solvable yet contains a large number of fortuitous states. We investigate these states exactly at finite $N$ and, in the large‑$N$ limit, via a unitary matrix model. In particular, we develop a way to ``follow $N$'' in the unitary matrix integral and study how the answer of the integral depends sensitively on the value of $N$.
}

\begin{document}
\maketitle

\section{Introduction}

``Fortuity" has recently been proposed as a sharp feature of supersymmetric black hole microstates in AdS/CFT \cite{Chang:2024zqi}.\footnote{See also \cite{Kinney:2005ej,Chang:2013fba,Grant:2008sk,Chang:2022mjp,Chang:2023zqk,Choi:2022caq,Choi:2023znd,Budzik:2023vtr,Budzik:2023xbr} for important studies of $Q$-cohomologies in $\mathcal{N}=4$ SYM that led to the development of this idea and for a partial list of subsequent developments and generalizations in \cite{deMelloKoch:2024pcs,Chang:2024lxt,Chang:2025rqy,Gadde:2025yoa,Hughes:2025car,deMelloKoch:2025ngs,deMelloKoch:2025cec,Chang:2025mqp}.} Distinct from the microstates of horizonless geometries, which can be extrapolated to the $N\rightarrow\infty$ limit while remaining supersymmetric, the existence of fortuitous states relies on intrinsically finite-$N$ effects and thus they cannot be extrapolated to infinite $N$ while remaining BPS. This agrees with the broader expectation that the detailed properties of black hole microstates should depend sensitively on $N$ \cite{Schlenker:2022dyo} and serves as a useful playground to investigate such dependencies in detail. 

What is the minimal set of ingredients that could catalyze fortuity? Known holographic examples give the impression that fortuity only emerges after the system reaches  sufficient ``complexity".  For instance, in the 4d $\mathcal{N} = 4$ $\SU (N)$ SYM theory, there is strong evidence suggesting that fortuity only happens in the $1/16$-BPS sector, while it does not appear in the much simpler $1/2, 1/4$ and $1/8$-BPS sectors \cite{Chang:2013fba,Chang:2023ywj}. A truncation of the $1/16$-BPS sector - the so-called BMN sector, achieves fortuity with seven species of fields \cite{Choi:2023znd,Choi:2023vdm}. 
On a different front, toy models of fortuity have been proposed, with the simplest example being the $\mathcal{N}=2$ SUSY SYK model \cite{Fu:2016vas,Chang:2024lxt}. This model involves only  a single vector-like field $\psi_i$, but it is complex in a different way, in that it has random couplings.

In this article, we show that to achieve fortuity, one only needs arguably the minimal ingredient - a \emph{single} fermionic matrix.
 To be precise, we consider a quantum mechanical model that contains $N^2$ complex fermions, organized into an $N\times N$ matrix $\Psi = (\Psi_{ij})_{N\times N}$. The fermions satisfy anticommutation relations
\begin{equation}\label{anti}
	\{ \Psi_{ij}, \bar{\Psi}_{kl} \} = \delta_{il} \delta_{jk}, \quad \{ \Psi_{ij}, \Psi_{kl} \}  = \{ \bar{\Psi}_{ij} , \bar{\Psi}_{kl} \} = 0\,. 
\end{equation}
We will use $\Psi_{ij}$ to denote the  fermion creation operators, while viewing their Hermitian conjugates $\bar{\Psi}_{ji} \equiv \Psi_{ij}^\dagger$ as the annihilation operators. To avoid confusion, we will not use the notation $\Psi^\dagger$ from now on. The model is defined by writing down its supercharge $Q$ and its Hamiltonian is then determined through the $\mathcal{N}=2$ supersymmetric algebra\footnote{In \cite{Chang:2024lxt}, a different model with supercharge $Q = \tr{\Psi^2 \bar{\Psi}}$ was briefly considered. This model does not contain fortuitous states and is therefore uninteresting for our current consideration. We thank Chi-Ming Chang, Bik Soon Sia and Zhenbin Yang for prior discussions. }
\begin{equation}\label{Q}
	Q = \tr{\Psi^3} \,, \quad \quad H = \{Q , \bar{Q}\}\,.
\end{equation}

The model in (\ref{Q}) can be viewed as a special instance of the $\mathcal{N}=2$ SYK model, but with non-random and in fact highly sparse couplings.\footnote{In the SYK context, it is customary to study more general $p$-body interactions. Our model also comes with natural generalizations of the form $Q = \textrm{Tr}[\Psi^p], \, p\in \mathbb{Z}_{\textrm{odd}}$ but we have not studied them in detail. } For this reason, we do not need to repeat the formal definition of fortuity here, since it was readily discussed in \cite{Chang:2024lxt} for the generic SYK models. 

As we will show, the model (\ref{Q}) contains fortuitous states, but not in the singlet sector. If one insists on having fortuitous states in the singlet sector, this can be achieved with two matrices through an almost trivial generalization of the one-matrix model. For this reason, we will focus on the one-matrix case for the most part and only briefly comment on the two-matrix case in Section \ref{sec:discussion}.

The model we study in this article is exactly solvable. Therefore, perhaps not surprisingly, it is non-chaotic and non-holographic. Furthermore, the fortuitous states do not exhibit an important $R$-charge concentration property proposed in \cite{Chang:2024lxt}. Therefore, our model only captures some of the ``\emph{skeleton}" of fortuity in holographic theories, but not its ``\emph{meat}". 

Some readers might be discouraged by the existence of fortuitous states that are not black-hole-like. They need not be discouraged. A key aspect of fortuity is that one could view the fortuitous BPS states as originating from the non-BPS sector. In a holographic theory, the non-BPS high energy states are generically black hole microstates, and this naturally suggests that the fortuitous states carry similar black hole features, such as their strongly chaotic nature.\footnote{Numerical evidence for the continuation of the chaotic property of states entering the BPS subspace was discussed in the $\mathcal{N}=4$ SYM in \cite{Budzik:2023vtr} and in the SYK models in \cite{Chang:2024lxt}. This was also termed ``chaos invasion" in \cite{Chen:2024oqv}.} On the contrary, in our model, even the high-energy states are non-chaotic. In such cases, there is no reason to expect the fortuitous states to behave differently. It remains a well-motivated conjecture that in actual holographic theories, fortuitous states correspond to BPS black hole microstates, while the monotone states correspond to the microstates of horizonless geometries.

\begin{figure}[t!]
    \centering
 \includegraphics[scale = 0.45]
{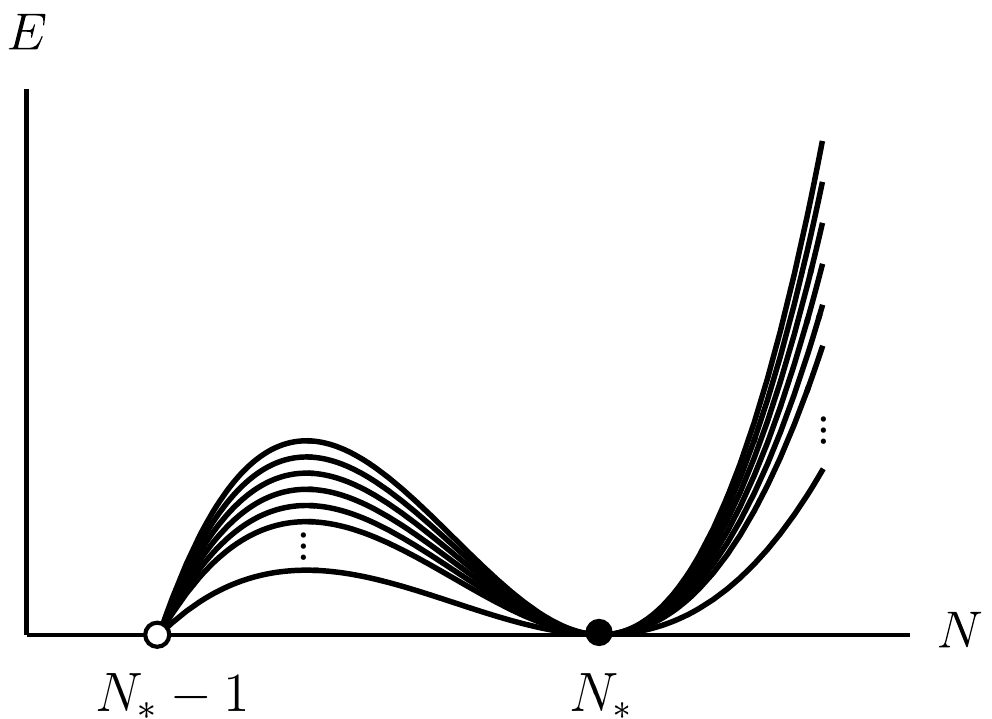}    
\caption{In our model, as well as in models such as the SUSY SYK model, the fortuitous states depend on $N$ in an extremely sensitive way. After an exponential number of states become BPS at $N= N_*$, they immediately become ``null" at $N = N_* - 1$.}
    \label{fig:sensitive}
\end{figure}

With that said, what is our motivation for studying this model further? The answer is that there are other interesting features about fortuity that are unrelated to chaos, and we can utilize the exactly solvable feature of our model to understand them better. One such feature is the extreme sensitivity of the fortuitous states to the precise integer values of $N$. In Figure \ref{fig:sensitive}, we illustrate the situation where an exponential (in $N^2$ for matrix theories) number of states enter the BPS subspace at a particular value of $N = N_*$, and then exit the Hilbert space by becoming null immediately at $N= N_* - 1$. Such extreme sensitivity, even though it has yet to be demonstrated clearly in the $\mathcal{N} = 4$ SYM theory, was seen in the SYK toy models and is expected more generally due to the $R$-charge concentration feature of fortuitous states. In our model, even though the $R$-charge concentration feature is not satisfied, we nonetheless see this sensitive dependence on $N$.\footnote{In a somewhat similar spirit, the $N$-dependence of the matrix elements of heavy operators in light states has been emphasized recently in \cite{Kudler-Flam:2025cki,Liu:2025cml} in relation to a puzzle about closed universes \cite{Antonini:2023hdh,Antonini:2024mci,Engelhardt:2025vsp}. In our case, the quantity that is sensitive to $N$ is the number of BPS states in a fixed charge sector.} 

We find this feature particularly puzzling from the perspectives of both the large $N$ expansion on the boundary side and the gravity picture. We are used to the idea that the exponentially many BPS or near-BPS states are described by a large $N$ saddle point, or in the bulk by a classical supersymmetric black hole geometry. What is the mechanism through which such a saddle point (or a classical geometry) hinges on the precise value of $N$ (or $G_N$), and manages to suddenly vanish if we decrease $N$ by one? Or perhaps there is a sense that the saddle point remains, but it ceases to contribute through some other mechanism?\footnote{In \cite{Chang:2024lxt}, this curious phenomenon was briefly commented upon and the corresponding saddle that ceases to contribute was phrased as a ``phantom black hole".}

In our model, we get to explore this sensitive dependence on $N$. In order to formulate the question clearly, we study a unitary matrix integral that captures the fortuitous states. It is familiar, from the study of large $N$ gauge theories, that the saddle points of the unitary matrix integral manage to capture some features of the bulk, even when the theory is free or weakly coupled \cite{Aharony:2003sx}.  It turns out that even though our model is non-holographic, the fortuitous states still manifest themselves in the unitary matrix integral as interesting saddle configurations of the eigenvalues. 
 We then devise a way to change $N$ by one in the matrix integral and track what happens to the saddle points. 
 What one finds is that the saddle configurations remain intact, but their contributions to the integral exactly cancel as $N$ is decreased by one.

The rest of this article is organized as follows. In Section \ref{sec:exact}, we will describe the exact spectrum of the model (\ref{Q}) and construct its fortuitous states explicitly. In Section \ref{sec:coll}, we study the sensitive dependence on $N$ of the fortuitous states in our model using the unitary matrix integral approach. 
We conclude with some further discussion in Section \ref{sec:discussion}. Some explicit computations are collected in the Appendices.

\section{Exact solution and fortuity}\label{sec:exact}

We can arrive at a different but elegant form of the supercharge in (\ref{Q}) by writing the fermion matrix $\Psi$ in a basis of $\mathfrak{su}(N)$ generators $T^a,\, a = 1,..., N^2 - 1$
\begin{equation}\label{adjbasis}
	\Psi = \frac{\psi^0}{\sqrt{N}} \mathbf{1} +  \sum_{a=1}^{N^2 - 1}  \sqrt{2}\,\psi^a T^a\,,
\end{equation}
where the Hermitian generators satisfy
\begin{equation}\label{alg}
	\textrm{Tr}[T^a T^b] = \frac{1}{2} \delta^{ab}, \quad [T^a , T^b] = \i f^{abc} T^c\,.
\end{equation}
It is straightforward to verify that the fermion operators $\psi^a,\, a = 1,...,N^2 - 1$ satisfy the standard anticommutation relation $\{\psi^a , \bar{\psi}^b\} = \delta^{ab}$ and $\{\psi^a , \psi^b\} =\{\bar{\psi}^a , \bar{\psi}^b\} = 0$. Using the adjoint basis (\ref{adjbasis}), we can rewrite the supercharge (\ref{Q}) as
\begin{equation}\label{Qnice}
\begin{aligned}
	Q & = 2 \sqrt{2 } \tr{T^a T^b T^c} \psi^a \psi^b \psi^c =\frac{\i }{\sqrt{2}}  f^{abc} \psi^a \psi^b \psi^c\,.
\end{aligned}
\end{equation}
where in the first equality, the trace mode $\psi^0$ decouples due to (\ref{alg}) and the fact that the fermion operators square to zero, while in the second equality, we used the fact that only the fully anti-symmetric part of $\tr{T^a T^b T^c}$, which is proportional to the structure constants, contributes. 
Since the trace mode $\psi^0$ decouples, its only role is to double the degeneracy of every energy level in the model where it is absent. We could therefore ignore it for most of the discussion and add it back in when necessary. 

The model has a $\textrm{U}(N)$ global symmetry, which contains a $\textrm{U}(1)$ factor whose charge is simply the fermion number 
\begin{equation}
	N_\Psi = \tr{\Psi \bar{\Psi}}\,.
\end{equation}
It also serves as the $R$-charge of the system, $[N_\Psi, Q] = 3 Q$. The fermions $\Psi_{ij}$ transform in the adjoint of the remaining $\textrm{SU}(N)$
\begin{equation}
	\Psi \, \rightarrow \, U \Psi U^\dagger, \quad U\in \textrm{SU}(N)\,.
\end{equation}
Crucially, we would like to consider the \emph{ungauged} model since, as it will soon be clear, fortuity does not occur in the singlet sector. We will discuss in Section \ref{sec:discussion} a generalization where fortuity happens within the singlet sector, and there one has the liberty to consider the gauged model without losing much. We can construct the generators of $\mathfrak{su}(N)$ as
\begin{equation}
	J^a = \tr{\Psi [T^a , \bar{\Psi}]} = -\i f^{abc} \psi^b \bar{\psi}^c\,, \quad \quad [J^a ,J^b] = \i f^{abc} J^c\,,
\end{equation}
and with them, the quadratic Casimir operator $\hat{C}_2$ is given by
\begin{equation}
	\hat{C}_{2} = \sum_{a=1}^{N^2 - 1}\, J^a J^a\,.
\end{equation}
With some fairly straightforward calculation, one finds that 
\begin{equation}\label{H}
	H = \{Q , \bar{Q}\} = 3 N(N^2 - 1) -9\,\hat{C}_2\,. 
\end{equation}
We see that once we decompose the Hilbert space into irreducible representations of $\textrm{SU}(N)$, the spectrum is completely determined  as it only depends on the quadratic Casimir of the representations. Therefore, the model is exactly solvable   and is non-chaotic.

To make our discussion less abstract, we provide the exact spectrum and the representations that appear for the cases of $N=2$ and $N=3$ in Table \ref{tab:degene}.

\begin{table}[t!]
\centering
\setlength{\tabcolsep}{8pt}
\renewcommand{\arraystretch}{1.15}

\begin{minipage}{0.35\linewidth}
\centering
\begin{tabular}{|c|cc|}
\hline
\multirow{2}{*}{$E$} & \multicolumn{2}{c|}{$N_\Psi$} \\
\cline{2-3}
 & 0 & 1 \\
\hline
18 & \textbf{1} &   \\
0  &   & \textbf{3} \\
\hline
\end{tabular}
\subcaption{$N=2$}
\end{minipage}\hfill
\begin{minipage}{0.60\linewidth}
\centering
\begin{tabular}{|c|ccccc|}
\hline
\multirow{2}{*}{$E$} & \multicolumn{5}{c|}{$N_\Psi$} \\
\cline{2-6}
 & 0 & 1 & 2 & 3 & 4 \\
\hline
72 & \textbf{1} &   &   & \textbf{1}   &   \\
45 &   &  \textbf{8} &  \textbf{8}  &  \textbf{8}  & $\textbf{8} \oplus \textbf{8}$ \\
18 &   &   & $\textbf{10} \oplus \overline{\textbf{10}}$    &  $\textbf{10} \oplus \overline{\textbf{10}}$      &   \\
0 &   &   &   &  \textbf{27} & $\textbf{27}\oplus \textbf{27}$  \\
\hline
\end{tabular}
\subcaption{$N=3$}
\end{minipage}
\caption{Energy levels and their corresponding irreducible representations in sectors with  fermion number $N_{\Psi}$, for $N=2$ and $N=3$. We only show here $N_{\Psi}< N^2/2$ as the table for  $N_{\Psi}\geq (N^2-1)/2$ simply follows by applying charge conjugation. We did not include the contribution from the decoupled $U(1)$ mode in the tables.}\label{tab:degene}
\end{table}

We note that the model has the interesting feature that the singlet states, which have $C_2 (\textbf{1}) = 0$, are the highest energy states. This is very different from holographic models, where singlet states are usually light. On the other hand, the lightest states, which turn out to have exactly zero energy, are those states with maximal possible eigenvalue of $\hat{C}_2$. We will colloquially call this representation the ``maximal representation" and denote it by $\r_*$.  We will now describe the maximal representation and some of its properties explicitly, including the fact that it is fortuitous.

\subsection{Maximal representation $\r_*$}\label{sec:rep}

We are interested in the representation with the maximal possible value of $C_2$ that one can get from multiplying the adjoint fermion creation operators $\Psi_{ij}$ and acting them on the Fock vacuum. Mathematically, this is equivalent to asking which $\SU (N)$ irreducible representations appear in the anti-symmetric product of $p$ copies of the adjoint representation
\begin{equation}\label{antiprod}
	\wedge^p  \textbf{adj}\,,
\end{equation} 
where $p=1,...,N^2$ corresponds to the fermion occupation number $N_{\Psi}$. 

This problem can be solved in various ways.\footnote{Similar problems also appear in the study of (1+1)-dimensional QCD with adjoint fermions, see for example \cite{Dempsey:2022uie}.} One way is through enumeration. Suppose we want to know whether a representation $\textbf{r}$ shows up in (\ref{antiprod}); then we can write down its group character $\chi_{\r} (U)$ as well as the character $\chi_{\wedge^p  \textbf{adj}}(U)$ and directly compute their inner product.  We will not proceed in this way here, even though it will play an important role later on in Section \ref{sec:coll}. What we will present below is a direct way to construct our desired irreducible representation, and we will find that it has the maximum $C_2$ by simply showing that the energy vanishes for such a representation and combining with the knowledge that the Hamiltonian is bounded below by zero by design.

Before we construct such a representation explicitly, we can first ask what one might expect. A representation with maximal $C_2$ should contain states built with a large number of fermion operators. By charge conjugation, we also do not expect the fermion number to be close to $N^2$ and the optimal value of $N_{\Psi}$ for a large representation would be around $N^2 /2$. 

We are therefore led to constructing a state that has roughly $N^2 /2$ fermion excitations, which would be a highest weight state that is annihilated by all the positive-root $\su (N)$ generators (namely, the analogue of the $J_+$ operator in the simplest $\su (2)$ case). A  convenient explicit basis of the positive-root generators is
\begin{equation}
	J_{E^{ij}} = \tr{\Psi [E^{ij}, \bar{\Psi}]}, \quad 1\leq i<j \leq N\,,
\end{equation}
where the matrix $E^{ij}$ is defined as
\begin{equation}
	(E^{ij})_{kl} = \delta^i_{k} \delta^{j}_l\,,
\end{equation}
namely, it has only a single non-zero element at the position indicated by the indices. The negative-root generators are  given by the same expressions but with $E^{ij}, i>j$, and finally, the Cartan generators are given by
\begin{equation}
	J_{H_i} = \tr{\Psi [H_i, \bar{\Psi}]}, \quad H_i \equiv E^{ii} - E^{i+1,i+1}, \quad i = 1,...,N-1\,.
\end{equation} 
Under the action of $J_{E^{ij}}$, the fermion creation operators transform as
\begin{equation}\label{JEtrans}
	[J_{E^{ij}} ,  \Psi_{kl}] = ([\Psi, E^{ij}])_{kl} = \delta_{jl} \Psi_{ki}   - \delta_{ki} \Psi_{ jl }\,
\end{equation}
while under $J_{H_i}$, we have
\begin{equation}
	[J_{H_i}, \Psi_{kl}] = \alpha_{kl}(H_i) \Psi_{kl}\,, \quad  \alpha_{kl}(H_i)  =  \delta_{li} - \delta_{l,i+1} - \delta_{ki}+\delta_{k,i+1} \,.
\end{equation}
\begin{figure}[t!]
\centering
\begin{tikzpicture}[>=Stealth, every node/.style={font=\large}]
  \matrix (m) [matrix of math nodes,
               nodes in empty cells,
               row sep=0.8em, column sep=0.8em,
               left delimiter=(, right delimiter=)]
  {
      & & & \\
      \star & & & \\
      \star & \star & & \\
      \star & \star & \star & \\
  };

  \draw[->, thick, bend left=40]
      (0.5,-1.5) to node[below, sloped, pos=.5] {$J_{E^{23}}$} (-0.3,-1.5);

\end{tikzpicture}
\caption{We can form a highest weight state $\ket{\lambda}$ by occupying all the fermion modes in the lower triangle. The figure illustrates the case when $N=4$. The arrow shows the action of a positive-root generator $J_{E^{23}}$ on  $\Psi_{43}$. It transforms $\Psi_{43}$ into $\Psi_{42}$, which is already occupied, and thus annihilating the state.}
\label{fig:highest}
\end{figure}
In this convention, all the fermion operators in the lower (upper) triangle  carry positive (negative) roots, while the traceless diagonal combinations of fermions form the Cartan of $\su (N)$ and carry zero weight. Therefore, if we consider a state $\ket{\lambda}$ where all the fermion modes in the lower triangle are occupied
\begin{equation}\label{high}
	\ket{\lambda} = \prod_{ i>j} \Psi_{ij} \ket{0}\,,
\end{equation}
it will reach the highest weight possible since all the positive-root fermions are occupied (and they can be occupied at most once due to the Pauli exclusion principle). $\ket{\lambda}$ is thus a highest weight state, with maximal possible weight $\lambda$ given by 
\begin{equation}\label{weight}
	\lambda = \sum_{i>j} \alpha_{ij} =   \underbrace{(2,2,...,2)}_\text{$N-1$} \,. 
\end{equation}
We can check explicitly that $\ket{\lambda}$ is highest weight by noting that according to the transformation rule (\ref{JEtrans}), when we commute any positive-root generators with $\Psi_{ij}, i>j$, we always get a different fermion mode that is also in the lower triangle, which is already occupied. Starting from the highest weight state in (\ref{high}), one can then straightforwardly build the entire representation by acting with the negative-root generators.

Apart from  (\ref{high}), one can consider a more general family of highest weight states by decorating (\ref{high}) with an arbitrary number of fermion modes in the Cartan.  In other words, the following states
\begin{equation}
	\ket{\tilde{\lambda}} = \prod_{\textrm{any}\,\, i} \Psi_{ii}\prod_{ i>j} \Psi_{ij} \ket{0}\,,
\end{equation}
are all valid highest weight states with the same weight  as in (\ref{weight}). 



\begin{figure}[t!]
    \centering

\def\N{8}            
\def\cell{0.4}      
\def\dRadius{0.04}  
\def\dSep{0.50}      
\def\dYOffset{0.5}  

\begin{tikzpicture}[x=\cell cm,y=\cell cm, every node/.style={font=\small}]
  \pgfmathtruncatemacro{\H}{\N - 1}     
  \pgfmathtruncatemacro{\W}{2*\N-2}   

  \fill[gray!12]
    (0,0) -- (0,\H) -- (\W,\H)
    \foreach \i in {1,...,\H}{ -- ++(0,-1) -- ++(-2,0) } -- cycle;

  \draw[line width=1.05pt,line cap=round,line join=round]
    (0,0) -- (0,\H) -- (\W,\H)
    \foreach \i in {1,...,\H}{ -- ++(0,-1) -- ++(-2,0) } -- cycle;

  \draw[decorate,decoration={brace,amplitude=6pt},thick]
    (0,\H+0.8) -- (\W,\H+0.8)
    node[midway,yshift=15pt] {$2N-2$};

  \draw[decorate,decoration={brace,amplitude=6pt,mirror},thick]
    (-0.85,\H) -- (-0.85,0)
    node[midway,xshift=-15pt,rotate=90] {$N-1$};

\end{tikzpicture}

    \caption{The Young diagram of the maximal representation $\r_*$. The $i$-th row has $2N-2i$ boxes and it has the maximum number of $N-1$ rows.}
    \label{fig:Young}
\end{figure}
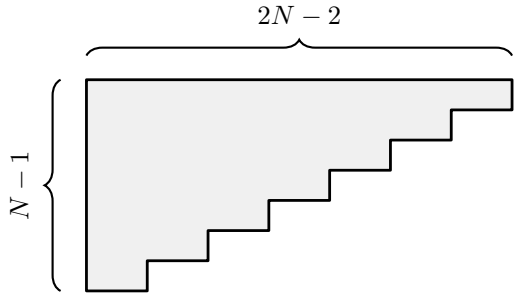

Using the weight of $\r_*$, it is straightforward to work out the Young diagram associated with the representation, which is shown in Figure \ref{fig:Young}. The Young diagram has a staircase shape and saturates the possible number of rows, i.e., $N-1$. This means that it is on the verge of being disallowed, a point that we will elaborate more on in the next subsection.

Given the Young diagram, it is a straightforward exercise to compute the quadratic Casimir as well as the dimension of the representation. We include this computation in Appendix \ref{app:group}. We find that
\begin{equation}
	C_{2}(\r_*) = \frac{N(N^2 - 1)}{3},
\end{equation}
which indeed exactly cancels out the constant part in the Hamiltonian in (\ref{H}) and leads to zero energy. Since the spectrum of the Hamiltonian is bounded below by zero, this also verifies indirectly our claim that we have found the representation with the largest $C_2$. The dimension of the representation is given by
\begin{equation}
	\textrm{dim} (\r_*) = 3^{\frac{N(N-1)}{2}}\,.
\end{equation}
We see that there is an order $N^2$ entropy in the BPS (zero energy) subspace. Of course, this is only due to the fact that we have a very large representation. The multiplicity of the representation does not grow as $e^{N^2}$.

With these, it is straightforward to write down the BPS partition function for our model, which encodes the number of BPS states with different fermion numbers
\begin{equation}\label{ZBPS}
	Z_{\textrm{BPS}} (q) \equiv \textrm{Tr}_{E=0}[q^{N_\Psi}] =   3^{\frac{N(N-1)}{2}} \times (1+q)^N q^{\frac{N(N-1)}{2}}\,.
\end{equation}
For later convenience, we also write down the partition function for $\r_*$, which simply encodes the multiplicity of $\r_*$ for a given $N_\Psi$
\begin{equation}\label{Zrs}
	Z_{\r_*} (q) = (1+q)^N q^{\frac{N(N-1)}{2}}\,.
\end{equation}
From (\ref{ZBPS}) or (\ref{Zrs}), we see that the BPS states exist when 
\begin{equation}\label{range}
	\frac{N(N-1)}{2} \leq N_{\Psi} \leq \frac{N(N+1)}{2}
\end{equation}
but not outside this range. We emphasize that this does \emph{not} satisfy the $R$-charge concentration phenomenon discussed in \cite{Chang:2024lxt}, which demands that within an irreducible cochain complex there is only a single cochain containing an exponential amount of states. For comparison, if we consider the $\mathcal{N}=2$ SUSY SYK model with $N^2$ fermions, then with generic couplings, the BPS states exist only within $ \frac{N^2}{2} -1 \leq N_{\Psi} \leq \frac{N^2}{2} +1 $.\footnote{Assuming $N$ is even. When $N$ is odd, the concentration is similarly sharp but the precise range is different, see \cite{Fu:2016vas,Kanazawa:2017dpd}.} In our model, the $\SU (N)$ symmetry constrains the interaction so severely that not all the states that can be lifted are lifted, resulting in a non-concentrated distribution.

\subsection{Fortuity of $\r_*$}

In Section \ref{sec:rep}, we have explicitly described the representation $\r_*$ that hosts all the BPS states. Now we explain that all such states are fortuitous. 

The quickest way to understand their fortuity is to observe that their fermion numbers are restricted in a range that depends explicitly on $N$, as in (\ref{range}). Since the fermion number $N_{\Psi}$ is invariant under the projection map used in the definition of fortuity \cite{Chang:2024lxt}, BPS states with any given $N_{\Psi}$ will inevitably be lifted at large $N$ as they are out of the range where BPS states are allowed, and are therefore fortuitous.

Using our explicit construction of the representation $\r_*$, we can give a more detailed characterization of fortuity. In particular, a nice way to ``see" fortuity is through the Young diagram. Here, the Young diagram plays a  role similar to the multi-trace expressions in models where the $\SU (N)$ symmetry is gauged, in that we can talk about Young diagrams by specifying their shapes without referring to the value of $N$ explicitly. This provides a way to interpolate between theories with different $N$ in a uniform way.

To illustrate this idea, we can consider a staircase Young diagram, with $2N_* - 2i$ boxes on the $i$-th row. Here, $N_*$ is just an arbitrary positive integer, used to specify the shape of the Young diagram. Now, we can embed this Young diagram in theories with different $N$, and depending on the relation between $N$ and $N_*$, the Young diagram can have very different meanings.

\begin{figure}[t!]
  \centering

  \begin{subfigure}[b]{0.3\linewidth}
    \centering

    \def\N{8}            
\def\cell{0.25}      
\def\dRadius{0.04}  
\def\dSep{0.50}      
\def\dYOffset{0.5}  

\begin{tikzpicture}[x=\cell cm,y=\cell cm, every node/.style={font=\small}]  
  \pgfmathtruncatemacro{\H}{\N - 1}     
  \pgfmathtruncatemacro{\W}{2*\N-2}   

 \fill[cyan, fill opacity=0.15] (0,-1) rectangle (\W + 2,\H);

  \fill[gray!12]
    (0,0) -- (0,\H ) -- (\W,\H )
    \foreach \i in {1,...,\H}{ -- ++(0,-1) -- ++(-2,0) } -- cycle;

  \draw[line width=1.05pt,line cap=round,line join=round]
    (0,0) -- (0,\H) -- (\W,\H)
    \foreach \i in {1,...,\H}{ -- ++(0,-1) -- ++(-2,0) } -- cycle;

  \draw[decorate,decoration={brace,amplitude=6pt},thick]
    (0,\H+0.8) -- (\W+2,\H+0.8)
    node[midway,yshift=15pt] {$2N-2$};

  \draw[decorate,decoration={brace,amplitude=6pt,mirror},thick]
    (-0.85,\H) -- (-0.85,-1)
    node[midway,xshift=-15pt,rotate=90] {$N-1$};
\end{tikzpicture}

    \caption{}
    \label{fig:Nbig}
  \end{subfigure}
  \hfill
  \begin{subfigure}[b]{0.3\linewidth}
    \centering
   
   \def\N{8}            
\def\cell{0.25}      
\def\dRadius{0.04}  
\def\dSep{0.50}      
\def\dYOffset{0.5}  

\begin{tikzpicture}[x=\cell cm,y=\cell cm, every node/.style={font=\small}]
  \pgfmathtruncatemacro{\H}{\N - 1}     
  \pgfmathtruncatemacro{\W}{2*\N-2}   

 \fill[cyan, fill opacity=0.15] (0,0) rectangle (\W ,\H);
  \fill[gray!12]
    (0,0) -- (0,\H) -- (\W,\H)
    \foreach \i in {1,...,\H}{ -- ++(0,-1) -- ++(-2,0) } -- cycle;

  \draw[line width=1.05pt,line cap=round,line join=round]
    (0,0) -- (0,\H) -- (\W,\H)
    \foreach \i in {1,...,\H}{ -- ++(0,-1) -- ++(-2,0) } -- cycle;

  \draw[decorate,decoration={brace,amplitude=6pt},thick]
    (0,\H+0.8) -- (\W,\H+0.8)
    node[midway,yshift=15pt] {$2N-2$};

  \draw[decorate,decoration={brace,amplitude=6pt,mirror},thick]
    (-0.85,\H) -- (-0.85,0)
    node[midway,xshift=-15pt,rotate=90] {$N-1$};

\path (0,\H+0.8) -- (\W+2,\H+0.8);
\path (-0.85,\H) -- (-0.85,-1);

\end{tikzpicture}

    \caption{}
    \label{fig:Nequal}
  \end{subfigure}
 \hfill
  \begin{subfigure}[b]{0.3\linewidth}
    \centering
   
   \def\N{8}            
\def\cell{0.25}      
\def\dRadius{0.04}  
\def\dSep{0.50}      
\def\dYOffset{0.5}  

\begin{tikzpicture}[x=\cell cm,y=\cell cm, every node/.style={font=\small}]
  \pgfmathtruncatemacro{\H}{\N - 1}     
  \pgfmathtruncatemacro{\W}{2*\N-2}   

   \fill[cyan, fill opacity=0.15] (0,1) rectangle (\W - 2,\H);
  \fill[gray!12]
    (0,0) -- (0,\H) -- (\W,\H)
    \foreach \i in {1,...,\H}{ -- ++(0,-1) -- ++(-2,0) } -- cycle;

  \draw[line width=1.05pt,line cap=round,line join=round]
    (0,0) -- (0,\H) -- (\W,\H)
    \foreach \i in {1,...,\H}{ -- ++(0,-1) -- ++(-2,0) } -- cycle;

  \draw[decorate,decoration={brace,amplitude=6pt},thick]
    (0,\H+0.8) -- (\W-2,\H+0.8)
    node[midway,yshift=15pt] {$2N-2$};

  \draw[decorate,decoration={brace,amplitude=6pt,mirror},thick]
    (-0.85,\H) -- (-0.85,1)
    
    node[midway,xshift=-15pt,rotate=90] {$N-1$};

  \path (0,\H+0.8) -- (\W+2,\H+0.8);
\path (-0.85,\H) -- (-0.85,-1);
  
\end{tikzpicture}

    \caption{}
    \label{fig:Nsmall}
  \end{subfigure}

  \caption{We fix the number of rows of the staircase Young diagram as $N_*$ and compare it with the value of $N$.  See main text for discussion. }
  \label{fig:fortuity}
\end{figure}
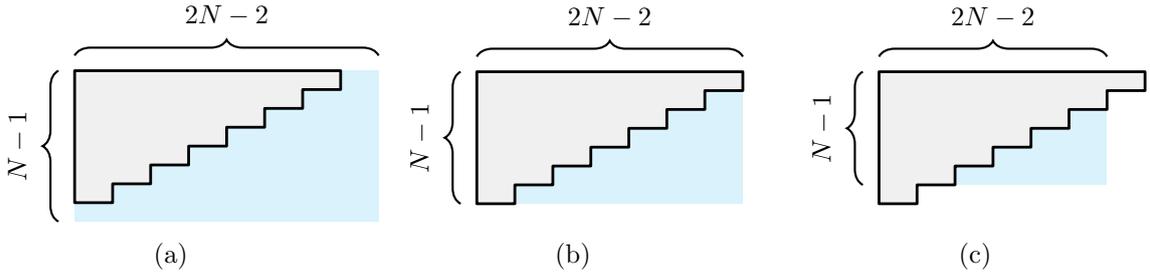

In Figure \ref{fig:fortuity}, we depict the three possible situations, $N > N_*$, $N= N_*$ or $N < N_*$. When $N > N_*$, shown in Figure \ref{fig:fortuity} (a), the Young diagram corresponds to a representation with weight
\begin{equation}
	 \overbrace{  ( \underbrace{2,...,2}_{N_* - 1} ,0,...,0 ) }^{N-1} 
\end{equation}
and in the convention of Section \ref{sec:rep}, it corresponds to a representation whose highest-weight state does not occupy all the lower-triangle (positive-root) fermions. Such states with non-maximal weight are not BPS as we found through our exact solution.

When $N= N_*$, as shown in Figure \ref{fig:fortuity} (b), we land on our discussion in Section \ref{sec:rep}. It is easy to see that now the highest weight state becomes annihilated by $Q$ since every term in $Q = \tr{\Psi^3}$ involves at least one fermion creation operator in the lower triangle.

Finally, when $N < N_*$, even by just one, the number of rows of the Young diagram exceeds the maximal allowed range for $\SU (N)$ and is no longer allowed. This is illustrated in Figure \ref{fig:fortuity} (c). Similar effects of Young diagram exceeding the allowed range also play an important role in the discussion of giant gravitons in AdS$_5 \times S^5$ \cite{McGreevy:2000cw,Corley:2001zk}.

\section{Large $N$ unitary matrix model analysis}\label{sec:coll}

As advocated in the introduction, we would like to use our model as a playground to explore fortuity using large $N$ techniques that are adaptable to other theories. 
To motivate our discussion, it is suggestive to rewrite our exact answer for $Z_{\r_*}(q)$ as 
\begin{equation}\label{action}
	Z_{\r_*} (q) = (1+q)^N q^{\frac{N(N-1)}{2}} \approx e^{- \frac{N^2}{2} \log \frac{1}{q}}\,.
\end{equation}
The fact that the expression has a non-zero ``classical action" of order $N^2$ hints that it may admit a saddle point interpretation.

What large-$N$ integral does the classical action in (\ref{action}) arise from?
 A natural candidate is the unitary matrix integral, which computes the multiplicities of various representations in the Hilbert space.  More specifically, one can compute the multiplicity of the representation $\r_*$ by the integral
\begin{equation}\label{UMI}
	Z_{\r_*} (q) = \int \mathcal{D} U\, \chi_{\r_*} (U) \exp\left[ \sum_{n=1}^{\infty} \frac{(-1)^{n+1} q^n}{n} \textrm{Tr}(U^n)\textrm{Tr}(U^{\dagger n})  \right],
\end{equation} 
where $\chi_{\r_*} (U) $ is the $\SU (N)$ group character for $\r_*$, given by
\begin{equation}\label{char}
	\chi_{\r_*} (U) = \prod_{1\leq i < j \leq N} \left(1 + 2 \cos (\theta_i - \theta_j)\right)
\end{equation}
where $e^{i\theta_i}, i = 1, ..., N$ are the eigenvalues of the holonomy matrix $U$. 
We explain the derivation of $\chi_{\r_*} (U)$ in Appendix \ref{app:group}. In (\ref{UMI}), $\mathcal{D}U$ denotes the Haar measure of $\SU(N)$, whose explicit form will be given shortly.

Our discussion of the integral (\ref{UMI}) will be divided into two parts. First, in Section \ref{sec:saddle}, we will study the large-$N$ saddle point(s) of the integral, which can be associated with the fortuitous states. It turns out that the saddle point problem is quite subtle, mostly due to the fact that 
(\ref{char}) is not positive-definite. We will nonetheless provide some limited understanding of the saddle points in some special limits. Then, in Section \ref{sec:followN}, we discuss in what way the saddle points are ``fortuitous". We propose a prescription in which we keep the representation fixed and vary $N$ in the unitary matrix integral. Essentially, we track the multiplicity of representation $\r_*$ while introducing a factor $\lambda \in [0,1]$ that penalizes the inclusion of the fermions from the last row and column.   This leads to a one-parameter family of unitary matrix models in which the potential for the last eigenvalue $\theta_N$ depends on $\lambda$. Our focus will be on studying the $\theta_N$
integral carefully and examine its consequences.

%

A natural idea to study the unitary matrix integral in (\ref{UMI}) is to exponentiate $\chi_{\r_*} (U)$ and view it as part of the classical action. In other words, we can introduce a continuous density of eigenvalues $\rho(\theta)$, satisfying $\int_0^{2\pi} \rho(\theta) \d\theta = 1$, and write
\begin{equation}\label{exponentiate}
\begin{aligned}
	\chi_{\r_*} (U) & = \exp\left[ \sum_{i<j} \log (1+ 2\cos(\theta_i - \theta_j))\right] \\
	& \approx  \exp\left[- N^2 \int \d\theta_1  \d\theta_2\,\rho(\theta_1)\rho(\theta_2) V_{\r_*}(\theta_1 - \theta_2)\right] \\
	& = \exp\left[-\frac{N^2}{2\pi}   V_{\r_*,0} -\frac{N^2}{\pi}  \sum_{n=1}^{\infty} |\rho_n|^2 V_{\r_*,n}\right] 
\end{aligned}
\end{equation}
where
\begin{equation}\label{log}
	V_{\r_*}(\theta) \equiv -\frac{1}{2}\log (1+ 2\cos(\theta)),   \quad V_{\r_*,n} \equiv \int_{-\pi}^\pi \d\theta\,	V_{\r_*}(\theta)  \cos (n\theta) 
\end{equation}
and we have expanded $\rho(\theta)$ as
\begin{equation}
	\rho(\theta) =\frac{1}{2\pi} + \frac{1}{\pi}\sum_{n=1}^\infty	 \rho_n \cos n\theta\,.
\end{equation}
One can then simply combine the Fourier coefficients $V_{\r_*,n}$ with the terms in the original action, and get 
\begin{equation}\label{Zrscoll}
	Z_{\r_*}  \approx \int \mathcal{D}U \exp \left[ -\frac{N^2}{2\pi}   V_{\r_*,0} + \sum_{n=1}^{\infty} \frac{a_n}{n}\textrm{Tr}(U^n)\textrm{Tr}(U^{\dagger n})  \right], \,\, a_n \equiv (-1)^{n+1} q^n - \frac{n}{\pi} V_{\r_*,n }\,.
\end{equation}
Such double trace matrix models can in principle be  solved exactly in the large $N$ limit, see \cite{Aharony:2003sx}. 
However, as readers might have already noticed, we've skipped over an important subtlety, that since the character $\chi_{\r_*}$ is not positive definite, we need to make sense of the logarithm factors in (\ref{exponentiate}) after the exponentiation. Suppose we fix the ambiguity of the logarithm by an $\i \epsilon$ prescription, i.e.,  view it as $\log (1+ 2\cos(\theta) + \i \epsilon)$ with $\epsilon > 0$, we can then compute the Fourier coefficients $V_{\r_*,n}$ and find that
\begin{equation}\label{complex}
	V_{\r_*,0} = - \frac{\i \pi^2}{3}, \quad V_{\r_*,n} = \frac{\pi}{n} e^{\i \frac{2\pi n}{3}}\,.
\end{equation}
In other words, we end up with a unitary matrix integral in which generically the coefficients $a_n$ are \emph{complex}, and as a consequence we also expect to find complex saddle points.

Having complex parameters and complex saddle points is not unheard of. Indeed, perhaps most famously, complex saddle points play an important role in the study of the superconformal index of the $1/16$-BPS sector in $\mathcal{N}=4$ SYM theory and are shown to match with bulk computation using supersymmetric black holes \cite{Cabo-Bizet:2018ehj,Choi:2018hmj,Benini:2018ywd}.\footnote{In the $\mathcal{N}=4$ SYM story, the complex fugacities arise (partly) due to the insertion of $(-1)^{F}$. We emphasize that here we are not computing an index and the complex coefficients have a different origin. } Nonetheless, the study of complex saddle points in our model appears to be challenging when using the method in \cite{Aharony:2003sx}, for reasons we will explain later. 
For this reason, we will retreat from the general problem (\ref{Zrscoll}) and consider special values of $q$ where things simplify dramatically. Only in Section \ref{sec:complex} will we provide some speculations for the picture at generic fugacity $q$. 

\subsection{Exact analysis for special values of the fugacity}\label{sec:saddle} 

A natural idea, familiar from the study of the superconformal index in $\mathcal{N}=4$ SYM, is to complexify the fugacity $q$ in such a way that we cancel  the oscillating signs in $\chi_{\r_*}$ as much as possible. It turns out that this cancellation can be achieved perfectly at $q = e^{\pm \i \frac{\pi}{3}}$, and up to a phase, the integrand in (\ref{Zrscoll}) becomes positive definite.\footnote{Experienced readers might be immediately worried that we are considering $|q|$ =1, which is often the location where one runs into various convergence issues. However, here we are considering a purely fermionic system; the Hilbert space is finite dimensional, and there is nothing wrong with considering $|q|=1$.}

To see why this is the case, it is more convenient to not exponentiate $\chi_{\r_*}$, but instead expand out the exponential in (\ref{UMI}). We have 
\begin{equation}\label{Zrsoriginal}
\begin{aligned}
	Z_{\r_*} (q) =\frac{(1+q)^N}{(2\pi)^N N!} \int \prod_{i=1}^N  \d \theta_i    \prod_{i<j} & \,\underbrace{ \left|e^{\i\theta_i} - e^{\i \theta_j}\right|^2 }_{\textrm{measure}}\underbrace{ \left(1+ 2 \cos \theta_{ij} \right) }_{\chi_{\r_*}}  \underbrace{ \left(1+ q  e^{\i \theta_{ij}}\right) }_{\Psi_{ij}} \underbrace{ \left(1+ q  e^{\i \theta_{ji}}\right) }_{\Psi_{ji}} 
\end{aligned}
\end{equation}
where $\theta_{ij} \equiv \theta_i - \theta_j$. Now, we observe that, if we set $q = e^{\pm \i \frac{\pi}{3}}$, we would have
\begin{equation}
(1+ q  e^{\i \theta_{ij}} ) ( 1+ q  e^{\i \theta_{ji}} )  = (1+q^2 + 2q \cos \theta_{ij}) = e^{\pm \i \frac{\pi}{3}} \left(1 + 2 \cos \theta_{ij}\right)
\end{equation}
and this nicely compensates the oscillating signs from the same factor in $\chi_{\r_*}$. In fact, the final integral has a surprisingly simple form (focusing on $q = e^{+ \i \frac{\pi}{3}}$ for simplicity)
\begin{equation}\label{Zr3theta}
\begin{aligned}
	Z_{\r_*}  & = e^{  \i \frac{N(N-1)}{2} \frac{\pi}{3} } \frac{(1+ e^{ \i \frac{\pi}{3}} )^N  }{(2\pi)^N N!}  \int \prod_{i=1}^N  \d \theta_i \prod_{i<j} \left|e^{\i\theta_i} - e^{\i \theta_j} \right|^2 (1+2 \cos \theta_{ij})^2  \\
	& =  e^{  \i \frac{N(N-1)}{2} \frac{\pi}{3} } \frac{(1+ e^{ \i \frac{\pi}{3}} )^N  }{(2\pi)^N N!}  \int \prod_{i=1}^N  \d \theta_i \prod_{i<j} \left|e^{\i 3\theta_i} - e^{\i 3\theta_j} \right|^2\,. 
\end{aligned} 
\end{equation}
We find that, up to some constant prefactors, the integrand becomes the familiar Vandermonde determinant expression, but with $\theta$ replaced by $3\theta$! 

\begin{figure}[t!]
    \centering
 \includegraphics[scale = 0.4]
{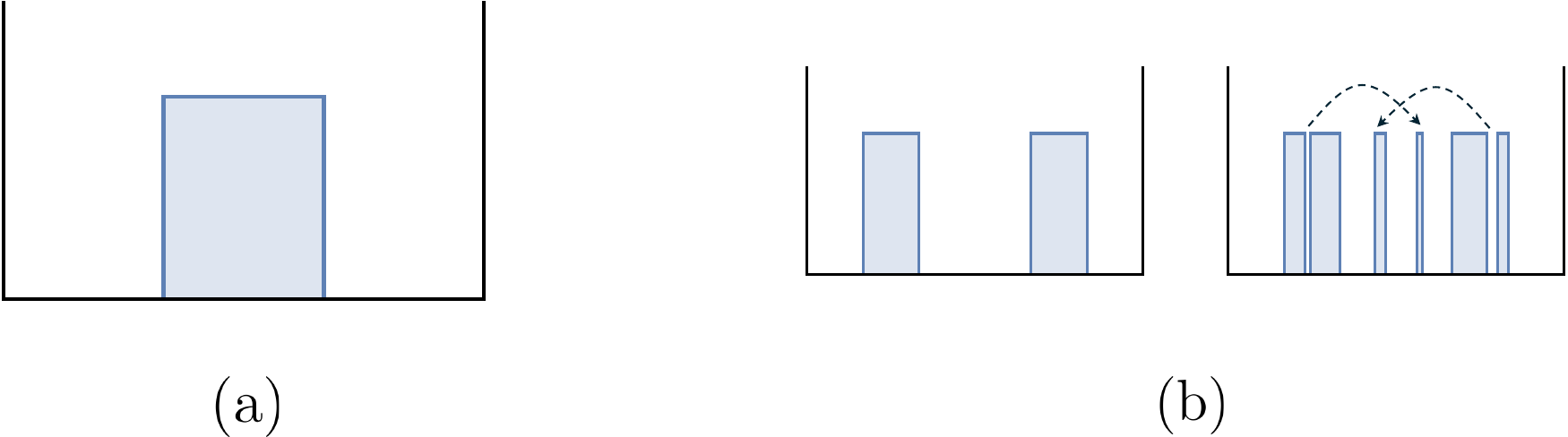}    
\caption{At special values $q =e^{\pm \i \frac{\pi}{3}}$, one can find a family of saddle points. This includes (a) a uniform distribution spanning an interval of length $2\pi/3$, as in (\ref{rhouniform}), but also includes multicut as well as highly fragmented saddle points shown in (b), arising by shifting blocks of eigenvalues by multiples of $2\pi/3$. }
    \label{fig:dist}
\end{figure}

The repulsive force between angle $3\theta_i , i = 1,..., N$ results in a saddle point where they are spaced evenly on the unit circle, which is
\begin{equation}\label{saddle3theta}
	3\theta_i = \frac{2\pi i}{N}, \quad i = 1,..., N\,,
\end{equation}
up to permutations. However, when we map the saddle configuration for $3\theta$ \label{saddle3theta} back to $\theta$, we do not get a single saddle point  but rather  a family of saddle points where each $\theta_i$ can be shifted by $\alpha_i \times  2\pi/3, \, \alpha_i = 0,1,2$, since $3\theta_i$ remains invariant under such shifts. In other words, we have
\begin{equation}\label{saddle}
	\theta_i = \frac{2\pi i}{3N} +\frac{2\pi}{3} \alpha_{i}, \quad \alpha_i = 0 ,1,2 , \quad i = 1,..., N\,.
\end{equation}
Within this family of saddle points (see Figure \ref{fig:dist} for an illustration), the simplest one has the form of a uniform density  which, up to an overall $\textrm{U}(1)$ rotation, can be written as
\begin{equation}\label{rhouniform}
	\rho (\theta) = \frac{3}{2\pi}\, , \quad - \frac{\pi}{3} < \theta < \frac{\pi}{3}\,.
\end{equation}
However, (\ref{saddle}) also hosts solutions which are multi-cut or even highly fragmented ones that cannot be well described by a smooth function $\rho(\theta)$. All of these saddle points have a large $N$ action that is zero, which is most easily seen by recalling that they descend from the same uniform saddle, which has zero action \cite{Aharony:2003sx}. All such configurations contribute equally in (\ref{Zr3theta}). Therefore, the large $N$ estimate for $Z_{\r_*}$ comes solely from the phase factor we pulled out in front of (\ref{Zr3theta})\footnote{The number of saddle points only grows as $3^N$ so it does not affect the $\mathcal{O}(N^2)$ action.}
\begin{equation}
	Z_{\r_*} (e^{ \i \frac{\pi}{3}}) \approx  e^{  \i \frac{N^2}{2} \frac{\pi}{3} } 
\end{equation}
which indeed reproduces (\ref{action}) once we plug in  $q = e^{\i \frac{\pi}{3}}$.

Even though all the configurations (\ref{saddle}) are valid maxima of the integrand (\ref{Zr3theta}) (modulo an overall $\textrm{U}(1)$ rotation zero mode), the situation is more subtle as we pass to the continuum description. We can look at the action for the Fourier modes  $\{\rho_n\}$ of $\rho(\theta)$, which has the form
\begin{equation}\label{zeromodes}
	\frac{S [\rho]}{N^2} = \sum_{n=1}^{\infty} \frac{1}{n}(1-a_n) \rho_n^2  \,.
\end{equation}
We note that when $q= e^{ \i \frac{\pi}{3}}$, from (\ref{Zrscoll}) and (\ref{complex}), we have\footnote{For $q= e^{ -\i \frac{\pi}{3}}$, one needs to use the opposite $\i \epsilon$ prescription around (\ref{complex}) for this argument.}
\begin{equation}
	a_{3n -2} = a_{3 n -1} = 1,\quad a_{3n} = -2, \quad n\in \mathbb{Z}\,.
\end{equation}
Therefore, all the Fourier modes $\rho_{3n-2}$ and $\rho_{3n-1}$ are exactly \emph{massless}, while the Fourier modes $\rho_{3n}$ are massive. Indeed, one can easily convince oneself that the class of configurations in (\ref{saddle}) corresponds to having $\rho_{3n-2}$ and $\rho_{3n-1}$ activated freely, while setting all the $\rho_{3n}$ to be zero. These zero modes are emergent only  in the continuum limit as they are absent in the finite $N$ integral (\ref{Zr3theta}).  For this reason, we will still refer to the various configurations loosely as ``saddle points".

\subsection{Following $N$ in the unitary matrix integral}\label{sec:followN}

In this section, we will describe a way to follow $N$ in the unitary matrix integral. The idea of following $N$ is similar to that in \cite{Budzik:2023vtr,Chang:2024zqi,Chang:2024lxt}, though instead of  following individual microstates, we will follow their large $N$ counterparts in the unitary matrix model. Our main interest will be to see how the saddle points in Section \ref{sec:saddle} manage to cease contributing as we decrease $N$ by one. 

We will do this by gradually decoupling the last row and column of fermions by introducing a penalty factor ${\color{red}{\lambda}}$ in the integral, namely
\begin{equation}\label{ZrchangeN}
\begin{aligned}
	\tilde{Z}_{\r_*} (q,\lambda)  \equiv & \frac{(1+q)^{N-1} (1+ {\color{red}{\lambda}}  q)}{(2\pi)^N N!} \int \prod_{i=1}^N  \d \theta_i   \prod_{i<j}  \,\underbrace{ \left|e^{\i\theta_i} - e^{\i \theta_j}\right|^2 }_{\textrm{measure}} \underbrace{ \left(1+ 2 \cos \theta_{ij} \right) }_{\chi_{\r_*}}   \\
	 &  \times \prod_{1\leq i<j \leq N-1} \underbrace{ \left(1+ q  e^{\i \theta_{ij}}\right) }_{\Psi_{ij}} \underbrace{ \left(1+ q  e^{\i \theta_{ji}}\right) }_{\Psi_{ji}} \prod_{i=1}^{N-1}\,  \underbrace{ \left(1+ {\color{red}{\lambda}}   q  e^{\i \theta_{iN}}\right) }_{\Psi_{iN}} \underbrace{ \left(1+ {\color{red}{\lambda}}   q  e^{\i \theta_{Ni}}\right) }_{\Psi_{Ni}} \,.
\end{aligned}
\end{equation}
When $\lambda = 1$ we simply have our original integral as in (\ref{Zrsoriginal}), while when $\lambda$ goes to zero, where we no longer include the fermions $\Psi_{iN}$ and $\Psi_{Ni}$, we should expect to find zero. This is because with our knowledge in Section \ref{sec:exact}, we would not have been able to build the representation $\r_*$ without using the last column and row of fermions. However, the fact that the integral goes to zero at $\lambda=0$ is not obvious from the expression (\ref{ZrchangeN}), since $\theta_N$ still enters non-trivially through both the measure factor and the character $\chi_{\r_*}$. 

We can verify the $\lambda$ dependence for small values of $N$ by doing the integral in (\ref{ZrchangeN}) exactly.\footnote{Practically, it is more convenient to write the integral in terms of complex variables $z_i = e^{\i \theta_i}$ and apply the residue theorem.} For example, we have
\begin{equation}\label{smallN}
\begin{aligned}
	& N = 3\, : \quad\quad \lambda^2(1+\lambda q)(1+q)^2 q^3 \,, \\
	& N = 4 \, : \quad\quad \lambda^3 (1+\lambda q)(1+q)^3 q^6\,, \\
	& N = 5\, :\quad\quad  \lambda^4 (1+ \lambda q) (1+q)^4 q^{10}\,, \\
	& ......
\end{aligned}
\end{equation}
which indeed agree with the exact expression (\ref{Zrs}) at $\lambda = 1$ and vanish at $\lambda = 0$.

We would like to see how this happens from a large $N$ expansion point of view. To do this, we can 
view the integral as divided into two parts: the integral for $\theta_1 , ..., \theta_{N-1}$, for which we can first look for the saddle points; and then  the integral for an additional eigenvalue $\theta_N$, which probes the background produced by the other angles. The logic here is similar to the study of eigenvalue instantons in large $N$ matrix models where we let one or a few eigenvalues wander off the continuum density of eigenvalues and look for new saddle points, but the difference is that here we are also modifying the potential for the last eigenvalue $\theta_N$ by a parameter $\lambda$.

\begin{figure}[t!]
    \centering
 \includegraphics[scale = 0.75]
{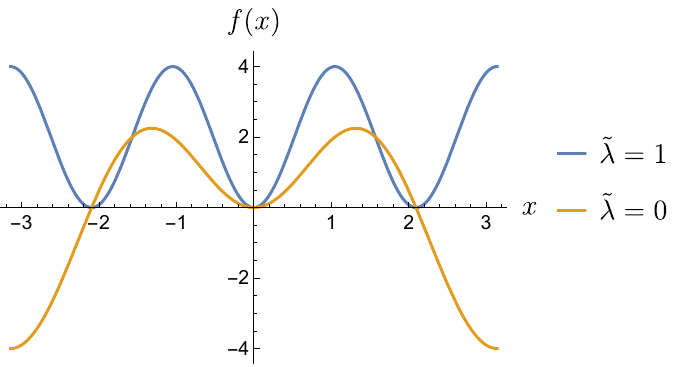}    
\caption{We plot the function $f$ in (\ref{deff}) through which the last eigenvalue $\theta_N$ enters the integral (\ref{Zintegral}). It is nonnegative when $\tilde{\lambda}=1$ but is no longer so when $\tilde{\lambda}=0$.}
    \label{fig:f}
\end{figure}

As a first pass, let us try to understand the situation at the special value $q=e^{\i \frac{\pi}{3}}$ where we have some control over the saddle point configurations. We have
\begin{equation}\label{Zintegral}
	\tilde{Z}_{\r_*} (q = e^{\i \frac{\pi}{3}},\lambda) \propto \int_{-\pi}^{\pi} \prod_{i=1}^{N-1} \d \theta_i \prod_{1\leq i<j \leq N-1} \left|e^{\i 3\theta_i} - e^{\i 3\theta_j}\right|^2 \times \int_{-\pi}^{\pi} \d \theta_N \prod_{i=1}^{N-1} f(\theta_{i N})
\end{equation}
where
\begin{equation}\label{deff}
	f(x) \equiv (2 - 2 \cos x  ) (1+2 \cos x) (1 + 2 \tilde{\lambda}\cos x ) , \quad \tilde{\lambda} \equiv\frac{  \lambda e^{\i \frac{\pi}{3} }  }{1 + \lambda^2 e^{\i \frac{2\pi}{3}} }\,.
\end{equation}
As shown in Figure \ref{fig:f}, when $\tilde{\lambda} = 1$, we have $f(x) \geq 0$. However, when $\tilde{\lambda} < 1$ it is no longer so, leaving the room for cancellations and the eventual vanishing of $\tilde{Z}_{\r_*}$ to happen. When $\tilde{\lambda}=0$, $f(x)$ is only nonnegative when $x \in [-2\pi/3, 2\pi/3]$.

Now, if we approximate the distribution of $\theta_1 ,... \theta_{N-1}$ by a continuous density $\rho(\theta)$, then we can write the integral over $\theta_N$ as\footnote{To properly define the integral, we can imagine placing $\theta_N$ slightly outside or inside the unit circle and then defining the potential through analytic continuation.}
\begin{equation}\label{lastint}
	 \int_{-\pi}^{\pi}  \d\theta_N \, \exp\left[ -N V_{\tilde{\lambda}}(\theta_N) \right]\,, \quad V_{\tilde{\lambda}}(\theta_N) = -\int_{-\pi}^{\pi} \d\theta \, \rho(\theta) \log  f(\theta - \theta_N)\,.
\end{equation}
Let us, for a moment, postpone the issue of the continuum zero modes and examine the integral (\ref{lastint}) for a given configuration of $\rho(\theta)$. 
As a simple example, we can look at the simplest configuration in (\ref{rhouniform}), with $\rho$ uniform between $-\pi/3$ and $\pi/3$, which gives
\begin{equation}\label{integral}
	V_{\tilde{\lambda}}(\theta_N) = - \frac{3}{2\pi} \int_{-\frac{\pi}{3}}^{\frac{\pi}{3}} \d\theta \,  \log  f(\theta - \theta_N)\,. 
\end{equation}
In Appendix \ref{app:integral}, we analyze this integral carefully. The conclusion is that, for small values of $\tilde{\lambda}$, the integral over $\theta_N$ is dominated by a saddle point at $\theta_N = \pi$, 
with action
\begin{equation}\label{Vlambda}
	V_{\tilde{\lambda}} (\pi) \approx    \i \pi  - \frac{2}{\pi} \textrm{Cl}_2 \left( \frac{\pi}{3} \right) + \frac{3\sqrt{3}}{\pi} \tilde{\lambda} + \mathcal{O} (\tilde{\lambda}^2)
\end{equation}
for small $\tilde{\lambda}$. Here $\textrm{Cl}_2$ is the Clausen function, with $\textrm{Cl}_2(\pi/3) \approx 1.0149$. 

\begin{figure}[t!]
  \centering
  \begin{subfigure}[b]{0.4\textwidth}
    \centering
   \begin{tikzpicture}
  \fill[blue!60, fill opacity=0.25]
    (0,0) -- (60:1.5cm) arc (60:300:1.5cm) -- cycle;

  \fill[red!70,  fill opacity=0.25]
    (0,0) -- (-60:1.5cm) arc (-60:60:1.5cm) -- cycle;

  \draw (0,0) circle (1.5cm);

  \newcommand{\NumDots}{10}
  \foreach \x in {1,2,...,\NumDots} {
    \pgfmathsetmacro{\Angle}{54 - 120/\NumDots*(\x-1)}
    \fill (\Angle:1.5cm) circle (1.5pt);
    \ifnum\x=1
      \node at (0.9,1.5) {$\theta_1$};
    \else
	\ifnum\x=\NumDots
          \node at (1.2,-1.55) {$\theta_{N-1}$};
      \fi
    \fi
  }
  \fill[red] (-1.5,0) circle (1.5pt);
  \node at (-1.85,0) {$\theta_N$};
\end{tikzpicture}
       \caption{}
    \label{fig:subfiga}
  \end{subfigure}
  \hspace{10pt} 
  \begin{subfigure}[b]{0.4\textwidth}
    \centering
    \begin{tikzpicture}
  \fill[blue!60, fill opacity=0.25]
    (0,0) -- (60:1.5cm) arc (60:300:1.5cm) -- cycle;

  \fill[red!70,  fill opacity=0.25]
    (0,0) -- (-60:1.5cm) arc (-60:60:1.5cm) -- cycle;

  \draw (0,0) circle (1.5cm);

  \newcommand{\NumDots}{5}
  \foreach \x in {1,2,...,\NumDots} {
    \pgfmathsetmacro{\Angle}{54 - 50/\NumDots*(\x-1)}
    \fill (\Angle:1.5cm) circle (1.5pt);
    \ifnum\x=1
      \node at (0.9,1.5) {$\theta_1$};
      \else
        \ifnum\x=\NumDots
          \node at (1.2,-1.55) {$\theta_{N-1}$};
      \fi
    \fi
  }
  \foreach \x in {1,2,...,\NumDots} {
    \pgfmathsetmacro{\Angle}{-10 - 50/\NumDots*(\x-1)}
    \fill (\Angle:1.5cm) circle (1.5pt);
    \ifnum\x=1
      \node at (0.9,1.5) {$\theta_1$};
      \else
        \ifnum\x=\NumDots
          \node at (1.2,-1.55) {$\theta_{N-1}$};
        \fi
    \fi
  }
  \fill[black] (-0.85,-1.24) circle (1.5pt);
  \node at (-1.0,-1.44) {$\theta_k$};
  \fill[red] (-1.5,0) circle (1.5pt);
  \node at (-1.85,0) {$\theta_N$};
\end{tikzpicture}
    \caption{}
    \label{fig:subfigb}
  \end{subfigure}
  \caption{(a) A configuration where the saddle point location of $\theta_N$ is on the other side of $\theta_1, ..., \theta_{N-1}$ on the unit circle, where they all satisfy $f(\theta_{iN})<0$  (represented by the red region). (b) We can also consider a configuration from (\ref{saddle}) where an angle $\theta_k$ is shifted by $\frac{4\pi}{3}$. We still have a saddle point for $\theta_N$ at approximately the same location but now $f(\theta_{kN})$ turns positive. This leads to a contribution that is similar in size but opposite in sign to the configuration in (a).}
  \label{fig:instanton}
\end{figure}
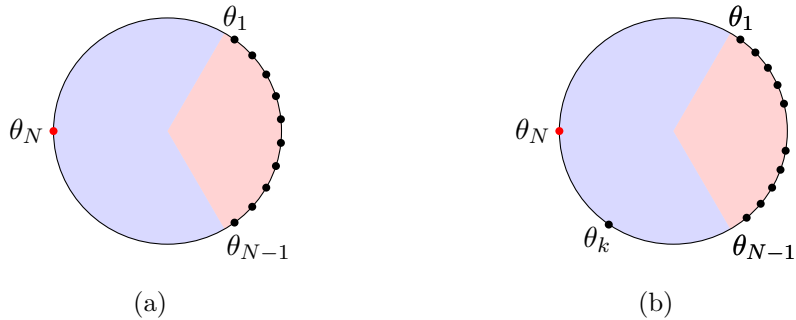

We illustrate this saddle point configuration in Figure \ref{fig:instanton} (a). The imaginary part in (\ref{Vlambda}) leads to a phase $e^{-\i \pi N}$, which has a clear interpretation in the figure. The eigenvalue $\theta_N$ is separated from all the rest of the eigenvalues by an angle that is greater than $2\pi/3$. Therefore, for $\tilde{\lambda} \ll 1$, we have
\begin{equation}
	f(\theta_{iN}) < 0 , \quad \quad  i = 1,..., N-1\,,
\end{equation}
so we expect a phase $(-1)^{N-1}$ in $\tilde{Z}_{\r_*}$, which agrees with $e^{-\i \pi N}$ within the precision of the large $N$ approximation. More importantly, the real part of (\ref{Vlambda}) is finite, meaning that the integral over $\theta_N$ simply gives a finite factor multiplying the contribution from the original saddle point.

The upshot of this example is to demonstrate that the integral over $\theta_N$ itself does not make the contribution from individual saddle points go to zero in the limit $\lambda \rightarrow 0$. 
Then, what is the mechanism through which $\tilde{Z}_{\r_*}(q, \lambda)$ vanishes as $\lambda \rightarrow 0$? At least for the special values of $q = e^{\pm \i \frac{\pi}{3}}$, it appears that the key ingredient is the cancellation among multiple saddle points. We recall that in the analysis leading to (\ref{Vlambda}), we ignored the fact that in the continuum, there are many zero modes in the fluctuations of $\rho(\theta)$, so fixing $\rho(\theta)$ and letting $\theta_N$ vary is not a good approximation in the continuum limit. In terms of the discrete eigenvalues, we need to sum over the set of configurations in (\ref{saddle}), and this leads to large cancellations.

To see how this cancellation happens, we notice that if we consider a different saddle configuration where a single $\theta_k$ is shifted by an angle $2\pi/3$ or $4\pi/3$, it will not change the saddle point location of the additional eigenvalue $\theta_N$ significantly. However, it will flip the contribution by an overall minus sign since now
\begin{equation}
	 f(\theta_{kN}) > 0 
\end{equation}
while we still have $f(\theta_{iN})<0$ for $i\neq k$. Therefore, the configuration in Figure \ref{fig:instanton} (b) comes with a relative minus sign compared to that in Figure \ref{fig:instanton} (a), leading to large cancellations. Of course, there is nothing preventing us from  shifting only one eigenvalue. Once we consider the full set of configurations in (\ref{saddle}), we expect a drastic cancellation among the contributions from various saddles.

To really see that such cancellations indeed lead to $\tilde{Z}_{\r_*}$ vanishing exactly at $\tilde{\lambda}$, we should however adopt a different perspective on the integral (\ref{Zintegral}) where we hold fixed the position of $\theta_N$ and integrate over the  rest of the eigenvalues $\theta_1, ..., \theta_{N-1}$. Here we are helped by the symmetry pattern of the integrand, where the part that does not involve $\theta_N$ is invariant under shifts of individual eigenvalues by multiples of $2\pi/3$. Therefore, we can average the part that is not invariant under the orbit of such shifts, and rewrite the integral (\ref{Zintegral}) as
\begin{equation}
	 \int_{-\pi}^{\pi} \prod_{i=1}^{N-1} \d \theta_i \prod_{1\leq i<j \leq N-1} \left|e^{\i 3\theta_i} - e^{\i 3\theta_j}\right|^2 \times  \int_{-\pi}^{\pi} \d \theta_N \prod_{i=1}^{N-1} \left[ \frac{1}{3} \sum_{\alpha_i =0}^{2} f\left(\theta_{i N} + \frac{2\pi}{3} \alpha_i \right) \right]\,.
\end{equation}
Using trigonometric identities, one finds that
\begin{equation}
	\frac{1}{3}\sum_{\alpha_i =0}^{2} f\left(\theta_{i N} + \frac{2\pi}{3} \alpha_i \right)  =\tilde{\lambda}  \left|e^{\i 3 \theta_i} - e^{\i 3 \theta_N}\right|^2\,.
\end{equation}
Therefore, we see that summing over such shifts makes the integrand vanish exactly when $\tilde{\lambda}\rightarrow 0$. Furthermore, as a byproduct, we also see that 
\begin{equation}
	\tilde{Z}_{\r_*} (q = e^{\i \frac{\pi}{3}},\lambda) \propto  \tilde{\lambda}^{N-1}  Z_{\r_*}(q = e^{\i \frac{\pi}{3}})
\end{equation}
reproducing the pattern found in (\ref{smallN}).

To summarize the discussion so far, we find that at $q = e^{\pm\i \frac{\pi}{3}}$, there is a large class of saddle point configurations for the eigenvalues. Their contributions sum up as long as we are sitting at the correct value of $N$. However, as we decrease $N$ by one in the matrix model by decoupling some of the fermions, we see that the configurations are dressed with varying signs and they cancel each other out exactly.

As a side remark, our prescription for following $N$ is not specific to the representation $\r_*$. We could have inserted characters for other representations, or simply insert $1$, which corresponds to the singlet sector, and ask similar questions. Of course, in more general cases, we wouldn't expect the integral to vanish at $\lambda = 0$. It would nonetheless be interesting to study more generally what happens as one follows $N$, as it might provide some hints on possibly generalizing the idea of fortuity beyond the BPS sector.

\subsection{Speculations about the picture at general fugacity}\label{sec:complex}

Much of our discussion so far is specific to the special values of the fugacity, i.e., $q = e^{\pm\i \frac{\pi}{3}}$. We have not been able to achieve a clear understanding of what happens for general $q$, but let's make some preliminary comments.\footnote{I'm grateful to Eunwoo Lee and Siyul Lee for helpful discussions of complex saddle points in unitary matrix integrals.}

When we move slightly away from the special point $q= e^{ \i \frac{\pi}{3}}$, the continuum zero modes $\rho_{3n-2}$ and $\rho_{3n-1}$ will be lifted. Depending on the direction one moves away from $q = e^{ \i \frac{\pi}{3}}$, the massless modes can either become tachyonic or massive.\footnote{Note that generically, the $a_n$ for the modes are complex; therefore, it is subtle to tell whether they are tachyonic or massive \cite{Copetti:2020dil}, which is usually determined by whether $a_n > 1$ or $a_n < 1$ when $a_n$ are real.} For example,
\begin{equation}
	\textrm{when} \quad q = e^{\i \frac{\pi}{3}} \pm 0.1\, , \quad a_1 = 1 \mp 0.1 \,, \quad  a_2 \approx  1 \pm (0.1  + 0.17 ) \,\i\, , \quad a_3 \approx -2  \, .
\end{equation}
The main challenge in finding the saddle point of this matrix model, using the method described in \cite{Aharony:2003sx} (see also \cite{Choi:2021lbk,Choi:2021rxi}) is that there are infinitely many modes becoming tachyonic at the same time.  
On the other hand, practically the method in \cite{Aharony:2003sx} requires truncating the matrix model to a finite number of terms $a_1 , ..., a_{n_{\textrm{max}}}$ and looking for convergence as $n_{\textrm{max}}$ increases. In our case, preliminary numerical exploration of this truncation does not seem to lead to results that suggest clear convergence as $n_\textrm{max}$ is taken to be large. 

It is a logical possibility that even away from the special values, the matrix model still admits a class of saddle points with the same action, and the mechanism through which the integral vanishes lies in the cancellation among them. In fact, the behavior of our matrix model near $q = e^{\i \frac{\pi}{3}}$ shares similar flavor to the one appearing in the discussion of ``small black holes" of AdS$_5 \times S^5$ in \cite{Choi:2021lbk}, where it was similarly suggested that there exists an infinite family of saddle points sharing the same large $N$ thermodynamics. Further work is required to illuminate such a phenomenon in matrix models.\footnote{The existence of a large number of soft modes in the collective description is also reminiscent of the quadratic SYK model \cite{Winer:2020mdc}. It is tempting to associate such behavior with the lack of chaos. I thank Steve Shenker for this interesting comment. }

However, 
it seems also plausible that when $q$ is moved away from the special values, there could be a single complex saddle point $\rho(\theta)$ that dominates and gives rise to the large-$N$ action (\ref{action}). If this were true, then what could be the mechanism through which the integral becomes zero as we decrease $N$ by one? It appears that instead of anything bad happening to the saddle point itself, it must be that the integral over the last eigenvalue $\theta_N$ becomes wildly oscillatory and gives exactly zero.

\section{Discussion}\label{sec:discussion}

In this article, we constructed and studied a simple supersymmetric matrix quantum mechanical model with a single complex adjoint fermion matrix. The model can be solved exactly using algebraic methods, and we find that there is a large number of fortuitous states within the irreducible representation $\r_*$ with the largest quadratic Casimir. 

Even though the model is non-chaotic, it retains interesting kinematic features of fortuity. In particular, similar to other models such as the SUSY SYK model \cite{Chang:2024lxt}, the fortuitous states in our model depend  on $N$ in an extremely sensitive way. Starting from the value of $N$ at which they become BPS, they all turn into ``null states" when $N$ is decreased by one. This can be understood pictorially in terms of the Young diagram for $\r_*$, which has the maximal number of rows and is no longer allowed if $N$ is decreased even slightly. 

We then proceed to understand this sensitivity in $N$ using the large $N$ unitary matrix integral. We devise a way to change the value of $N$ in the matrix model by one and ``watch" how the fortuitous states disappear in the matrix model. At the technical level, this involves separating the integration variables of the matrix model into $\theta_1 , ...,\theta_{N-1}$ and $\theta_N$, while tuning the dependence on $\theta_N$ in such a way that we interpolate between the $N$ and $N-1$ theories. At a purely pictorial level, one can think of the eigenvalues $\theta_1 , ... ,\theta_{N-1}$ as forming the bulk spacetime, while $\theta_N$ corresponds to an additional D-brane that is exploring the background.\footnote{Such a picture has also been studied in the context of the giant graviton expansion \cite{Gaiotto:2021xce}, see \cite{Murthy:2022ien,Chen:2024cvf}.}  

At certain special values of the fugacity, where we can understand the matrix model in greater detail, we find that the fortuitous states correspond not to a single but to a family of saddle point configurations of eigenvalues. As we decrease $N$ by one, the saddle point configurations of $\theta_1 ,... , \theta_{N-1}$ remain intact. However, the additional $\theta_N$ integral dresses them with sign-varying prefactors which lead to huge cancellation. In a sense, the original saddle points are still there in the theory with smaller $N$, but they become ``phantom" due to intricate cancellations \cite{Chang:2024lxt}.
It is important to stress, however, that for holographic theories, or possibly even for generic values of the fugacity in our model, one expects that there is a single (rather than many) saddle configuration to dominate. The natural  speculation is that the contribution from such a single saddle point vanishes upon integrating over the last eigenvalue $\theta_N$, but we have not demonstrated the details in such cases.

Below, we mention some further generalizations and future directions.

\paragraph{Following $N$ in more general unitary matrix integrals}
It is interesting to generalize some of the ideas presented here, especially that of following $N$ in the unitary matrix integral, to theories such as $\mathcal{N}=4$ SYM. We should emphasize that what we are doing is not as straightforward as simply changing $N$ in the matrix integral, which would lead to a smooth behavior in the large $N$ limit. This smooth behavior arises because, as we change $N$, we are also secretly focusing on fortuitous states in different charge sectors. Instead, what one really wants to do is, in some way, to fix the particular operators and track how they react to the change in $N$.  In our case, we are aided by the fact that the fortuitous states are in a specific representation, so we can focus on them by  keeping the group character fixed. In $\mathcal{N}=4$ SYM, where all operators are in the singlet sector, it is less obvious how to proceed. Perhaps one can utilize the restricted Schur polynomial techniques (see \cite{deMelloKoch:2024sdf} for a review), or draw some inspiration from other collective field techniques for huge operators in matrix models (see e.g. \cite{Berenstein:2008eg,Guerrieri:2025ytx,Kazakov:2024ald,Anempodistov:2025maj}).

\paragraph{Generalization with singlet fortuitous states}

The readers might be unsatisfied with the fact that there are no fortuitous states in the singlet sector of our model, which differs from more familiar examples such as $\mathcal{N}=4$ SYM. However, there is a  nearly trivial generalization of our model that would contain fortuitous states in the singlet sector. The idea is to take two independent fermionic matrices $\Psi_1$ and $\Psi_2$, each satisfying the anticommutation relation as in (\ref{anti}). We then take the supercharge to be
\begin{equation}
	Q = \tr{\Psi_1^3 } + \tr{\Psi_2^3}\,.
\end{equation}
$\Psi_1$ and $\Psi_2$ are completely decoupled and the Hamiltonian of the system is the sum of the Hamiltonian for individual matrices, both given by (\ref{H}). However, now if we consider the singlet sector, we obtain BPS states since they can arise in the tensor product of two copies of the representation $\r_*$, one from each of the fermion matrices 
\begin{equation}
	\textbf{1} \subset \r_{*, \Psi_1} \otimes \r_{*, \Psi_2}  \,.
\end{equation}
However, it should be noted that in this way we do not get an exponential number of singlet fortuitous states, nor do they satisfy $R$-charge concentration.

\paragraph{Simplest chaotic matrix model with $R$-charge concentration}

It is interesting to ask what is the simplest matrix quantum mechanical model that not only contains fortuitous states but also is chaotic and satisfies the $R$-charge concentration phenomenon. Such a model, in the large $N$ limit, will likely have the super-Schwarzian \cite{Stanford:2017thb} as its low energy effective theory. Within models with purely fermionic matrices and cubic supercharges (which can perhaps be called ``matrix SYK models"), it appears to us that the magic number might be three. For instance, as an example that is not particularly special but serves the purpose for illustration, a numerical simulation of the model with the following supercharge
\begin{equation}
	Q = \sum_{1\leq i\leq j\leq k\leq 3 }\tr{\Psi_i \Psi_j \Psi_k}
\end{equation}
at $N=2$ (thus $12$ complex fermions) suggests $R$-charge concentration. Other fermionic matrix models that are non-supersymmetric have also been studied in the literature, see \cite{Anninos:2015eji,Anninos:2016klf,Tierz:2017nvl,Klebanov:2018nfp,Gaitan:2020zbm} and references therein. It would be interesting to explore new techniques, such as the quantum mechanical matrix bootstrap (see e.g. \cite{Lin:2025srf,Laliberte:2025xvk}), to understand the properties of these models.

\vskip.5cm
 \noindent {\it {Acknowledgements}}  
We thank Chi-Ming Chang, Eunwoo Lee, Ji Hoon Lee, Siyul Lee, Henry Lin, Henry Maxfield, Stephen Shenker, Douglas Stanford, and Haifeng Tang for interesting discussions, and we also acknowledge useful conversations with ChatGPT. We thank Chi-Ming Chang for interesting comments on the draft. YC acknowledges support from DOE grant DE-SC0021085. Part of this work was performed during the Aspen summer program ``Recent developments in String Theory,'' supported by National Science Foundation grant PHY-2210452, where I also presented related results. YC also thanks the hospitality of Université Libre de Bruxelles and KU Leuven, where this work was presented.

\appendix

\section{Details of the representation $\r_*$}\label{app:group}

We have a Young diagram with $\lambda_j = 2N - 2j, \, j = 1 ,...,N-1$. 

We have the character formula
\begin{equation}
	\chi (U) = \frac{ \det \left[ x_{k}^{\lambda_j + N - j}\right]_{1\leq j,k \leq N}}{ \prod_{1\leq i<j \leq N} (x_i - x_j) }
\end{equation}
where $x_i =e^{i\theta_i}$, satisfying $\sum_i\theta_i = 0$. In our case, we have
\begin{equation}
	\det \left[ x_{k}^{\lambda_j + N - j}\right]_{1\leq j,k \leq N} = 	\det \left[ x_{k}^{2N - 2j + N - j}\right]_{1\leq j,k \leq N} = \det \left[ (x_{k}^3)^{N-j}\right]_{1\leq j,k \leq N}
\end{equation}
which is simply the Vandermonde determinant $\prod_{1\leq i < j\leq N} (x_i^3 - x_j^3)$. We then have
\begin{equation}
	\chi_{\r_*} (U) =\prod_{1\leq i < j\leq N}  \frac{x_i^3 - x_j^3}{x_i - x_j } = \prod_{1\leq i < j\leq N} ( x_i^2 + x_j^2 + x_i x_j )  = \prod_{1\leq i < j\leq N} ( 1 + 2 \cos (\theta_i - \theta_j ) ) 
\end{equation}
as claimed in (\ref{char}). Using the character, we immediately get
\begin{equation}
	\textrm{dim}(\r_*) = \chi_{\r_*} (1) = 3^{\frac{N(N-1)}{2}}\,.
\end{equation}
The quadratic Casimir is given in terms of the Young diagram shape by
\begin{equation}
	C_2 = \frac{1}{2} \left[ N \sum_{i} \lambda_i + n (\lambda_i - 2i + 1)  -  \frac{n^2}{N}\right] , \quad n = \sum_{i=1}^N \lambda_i \,.
\end{equation}
Specifying to our case, we have
\begin{equation}
\begin{aligned}
	C_{2} (\r_*) & = \frac{1}{2} \left[N \times N(N-1) + \frac{N(N-1)(2N-1)}{3} - \frac{(N(N-1))^2}{N}\right] \\
	& = \frac{N(N^2- 1)}{3}\,.
\end{aligned}
\end{equation}

\section{Integral over the last eigenvalue $\theta_N$}\label{app:integral}

In this appendix, we analyze the integral in (\ref{integral}) more carefully. We would like to start with a different integral where the range of the integral runs from $-\theta_0$ to $\theta_0$, with $\theta_0 < \pi/3$
\begin{equation}\label{intdiff}
	\int_{-\theta_0}^{\theta_0} \d \theta  \left[ \log (2 - 2 \cos (x-\theta) ) + \log (1  + 2 \cos (x-\theta) ) + \log (1 + 2 \tilde{\lambda} \cos (x- \theta)) \right]
\end{equation}
To ease the discussion, let's consider the case where $\tilde{\lambda} < 1/2$ so there is no branch cut issue from the third term in (\ref{intdiff}). The only term that requires extra attention is the second term. For every value of $\theta$, there is a branch cut running from $\theta + \frac{2\pi}{3}$ to $\theta + \frac{4\pi}{3}$. We define the $\log$ by giving $x$ a small imaginary part $x+ \i 0^+$. So when $x$ enters $[ \theta + \frac{2\pi}{3},\theta + \frac{4\pi}{3}]$, the imaginary part of $\log (1+ 2 \cos (x-\theta))$ goes from $0$ to $-\pi$, and when $x$ exits the cut, the imaginary part goes to $-2\pi$.

We want to show that $x=\pi$ is a stationary point of the integral and compute its action.
When $\theta_0< \pi/3$, one has that the imaginary part of (\ref{intdiff}) is a constant around $x = \pi$. This is simply because $1+ 2 \cos (\pi -  \theta) = 1 - 2 \cos \theta$ is negative for any $\theta \in [-\theta_0 , \theta_0]$.  Therefore, we only need to show that the real part of (\ref{intdiff}) is also stationary around $x=\pi$. This is clearly true simply due to the reflection symmetry of the integral under $x \rightarrow 2\pi - x$. 

Now, ultimately, we are interested in the case where $\theta_0 = \pi/3$ from (\ref{integral}) of our main text. Therefore, one might worry about the effect coming from the end points, where $1 + 2 \cos(\frac{\pi}{3}) = 0$.
To understand this properly, we have to go back to the finite $N$ picture, where we have $N-1$ eigenvalues in the cut and an extra one at $\pi$. If we do not have $\theta_N$, then $\theta_1, ..., \theta_{N-1}$ are spaced evenly from $- \pi/3$ to $\pi/3$ in the saddle point we are considering. However, once we introduce $\theta_N$ and place it at $\theta_N = \pi$, it gives rise to a force coming from the potential $-\log (1+2 \cos (\theta_1 - \theta_N) )$, which leads to a new balance where $\theta_1$ remains strictly inside $(- \pi/3, \pi/3)$ and similarly for $\theta_{N-1}$.\footnote{One can also reach equilibrium with them being outside the interval $(- \pi/3, \pi/3)$. This would correspond to other configurations such as in Figure \ref{fig:instanton} (b) and is not of our interest here.} Such backreaction effect can usually be ignored in the large $N$ limit but here it does play an important role. Therefore, there exists a saddle configuration where all $\theta_1 , ..., \theta_{N-1}$ are inside $-\pi/3$ to $\pi/3$ and can be described by our integral (\ref{intdiff}) with $\theta_0$ approaching $\pi/3$ from below.

\begin{figure}[t!]
    \centering
 \includegraphics[scale = 0.75]
{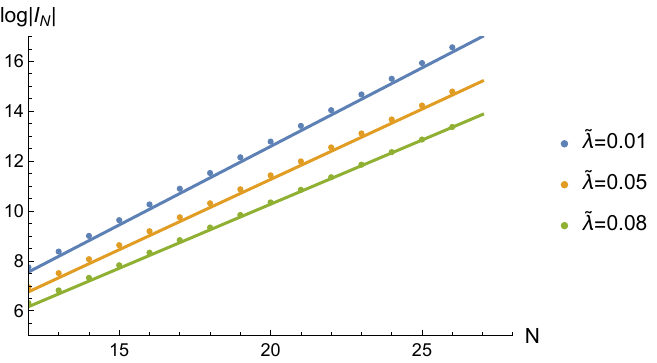}    
\caption{Comparison of the finite $N$ integral (\ref{IN}) with the large $N$ approximation (\ref{Vlambda}) for the configuration in (\ref{confapp}).}
    \label{fig:match}
\end{figure}

Computing the integral (\ref{intdiff}) with $\theta_0 = \pi/3$ and $\tilde{\lambda} \approx 0$, we find 
\begin{equation}\label{Vlambda}
	V_{\tilde{\lambda}} (\pi) \approx    \i \pi - \frac{2}{\pi} \textrm{Cl}_2 \left( \frac{\pi}{3} \right) + \frac{3\sqrt{3}}{\pi} \tilde{\lambda} + \mathcal{O} (\tilde{\lambda}^2)\,,
\end{equation}
 as quoted in the main text. Let's verify this large $N$ expression with finite $N$ numerics. Let us fix the location of the angles $\theta_1, ..., \theta_{N-1}$  to be
 \begin{equation}\label{confapp}
 	\theta_i = \frac{2\pi i}{N}, \quad i = 1,...,N-1\,.
 \end{equation}
We can then numerically perform the integral
\begin{equation}\label{IN}
	I_N = \int_{-\pi}^{\pi} \d \theta_N \prod_{i=1}^{N-1} f(\theta_{i N})\,.
\end{equation}
In Figure \ref{fig:match}, we compare the numerical integral with the large $N$ estimates and find a good match.

\bibliographystyle{JHEP}
\bibliography{references}
\end{document}